\documentclass[a4paper]{jpconf}
\usepackage{graphicx}
\begin{document}

%newnewnew---------------------------------------------------------************
\newcommand{\be}{\begin{equation}}
\newcommand{\ee}{\end{equation}}
\newcommand{\braket}[2]{\langle #1|#2\rangle}       % Inner Product
\newcommand*{\ket}[1]{|#1\rangle}
\newcommand*{\bra}[1]{\langle #1|}
\newcommand*{\cS}{\mathcal{S}}

\newcommand{\barr}{\begin{eqnarray}}
\newcommand{\earr}{\end{eqnarray}}
\newcommand{\ra}{\rangle}
\newcommand{\la}{\langle}
\newcommand{\beq}{\begin{equation}}
\newcommand{\eeq}{\end{equation}}
\newcommand{\Ud}{U^{\dagger}}
\newcommand{\up}{\uparrow_x}
\newcommand{\dn}{\downarrow_x}

\title{Weak measurement and weak values --- New insights and
effects in reflectivity and scattering processes}

\author{C  A  Chatzidimitriou-Dreismann}

\address{ Institute of Chemistry, Sekr.~C2, Faculty II, Technical University of Berlin,
 D-10623 Berlin, Germany}

\ead{dreismann@chem.tu-berlin.de}

\begin{abstract}
  Recently, the notions of Weak Measurement (WM), Weak Value (WV) and 
 Two-State-Vector Formalism (TSVF), firstly introduced by Aharonov and collaborators,
	have extended the theoretical frame of standard 
	quantum mechanics, thus providing 
 a quantum-theoretical formalism for extracting new information from a
system in the limit of small disturbance to its state. Here we provide an
application to the case of two-body scattering with one body  weakly 
interacting with its environment --- e.g. a neutron being scattered from 
a H$_2$ molecule physisorbed in a carbon nanotube. 
   In particular, we make contact with the field of incoherent inelastic neutron 
scattering from condensed systems.
We provide a physically compelling prediction of a new
quantum effect --- a momentum transfer deficit; or equivalently, an enhanced energy transfer; or 
an apparent  reduction of the mass of the   struck particle. 
E.g., when a neutron collides with a H$_2$
molecule in a C-nanotube and excites its translational motion along the nanotube, 
it apparently exchanges energy and momentum
with a fictitious particle with mass of 0.64 atomic mass units. 
Experimental results are shown and discussed in the new theoretical frame. 
The effect under consideration   has no conventional interpretation, thus also supporting the novelty 
of the quantum theoretical framework of WV and TVSF. Some speculative remarks about  possible applications 
being of technological interest (fuel cells and hydrogen storage;  Li$^+$ batteries; information and 
communication technology) are shortly mentioned.      
\end{abstract}

\section{Introduction}

  The fundamental $t$-inversion symmetry of the Schr\"odinger equation  plays a crucial role in the 
 novel theory of weak measurement (WM), weak values (WV) and the two-state vector formalism (TSVF) 
 of Aharonov and collaborators 
     \cite{Aharonov-Buch,AAV1988,Aharonov1990,Kofman2012,Tamir2013,Aharonov2014,Dressel2014}. 
 Based on this theory, new experiments were suggested and several novel quantum effects were discovered; 
 see the cited references.   
 
 Very recently, Aharonov et al.~\cite{Aharonov-NJP} provided a remarkably simple and clear example demonstrating the 
predictive power of the 
  theory, revealing an "anomalous"  momentum exchange  between photons (or particles) passing through a 
  Mach-Zehnder interferometer (MZI) and one of  its mirrors. Here the measured photon's 
(particle's) final state,  being post-selected in a specific output of the   MZI,  plays a crucial and succinct role,
together with the basic $t$-inversion  symmetry of quantum mechanics,
as captured by the Aharonov-Bergmann-Lebowitz rule \cite{ABL1964}. 
In essence, the revealed  effect (which has no conventional interpretation) is as follows: 
 Although the photons (particles) collide  with the considered mirror only from the inside of the MZI, they 
do not push the mirror outwards, but rather they somehow succeed to pull it in \cite{Aharonov-NJP}; see Sec.~2.

Inspired by this remarkable theoretical result, 
we  recently 
proposed an application of the theory under consideration, in the context of incoherent inelastic neutron 
scattering off atoms and/or molecules in condensed matter 
 \cite{Dreismann-Quanta}. Related  experimental results obtained by
 incoherent inelastic neutron scattering (IINS, or INS) and deep-inelastic neutron scattering (DINS) --- also called neutron
  Compton scattering (NCS) --- were presented and discussed  \cite{Dreismann-Quanta}.
 
   In this paper, after a short introduction to the new theory and the aforementioned "anomalous" 
	momentum transfer in a MZI, we present the specification of the theoretical frame to real 
	scattering experiments. In particular, we discuss a counter-intuitive experimental result of neutron
	scattering from H$_2$ in carbon nanotubes and related materials, as obtained with conventional INS as well as
	modern 2-\textit{dimensional}
	neutron spectrometers,   e.g. ARCS \cite{ARCS}. Additionally, an "anomalous" DINS result from H of a solid
	polymer is presented. 
	In short, our findings correspond to a striking  strong \textit{mass deficit} of the scattering objects. 
	These observations have no known conventional interpretation.
   
    The experimental findings and their theoretical interpretation support the view that the quantum theory of 
		WM, WV and TSVF sheds new light on interpretational issues concerning fundamental quantum theory. Moreover, 
this $t$-symmetric theoretical formalism  offers a new guide for our intuition to design  new experiments and discover 
 novel quantum effects. 

 \section{\label{NJPpaper} Motivation---A recent theoretical result of quantum optics}

The experimental context of this paper  concerns\\ 
(a) the measurements of \emph{momentum} and \emph{energy} transfer
in real scattering experiments, \\
(b) the predictions of conventional (classical or quantum) theory and\\ 
(c) comparison of the experimental results 
with a new
prediction based on the formalism of WM and TSVF.\\
 The latter point may be best motivated by
referring  directly to some of the intriguing results presented in a recent theoretical paper by 
Aharonor et al. \cite{Aharonov-NJP}.

Let us first refer to a surprising theoretical prediction derived by Aharonov et al. in
Ref.~\cite{Aharonov-NJP}. Here we consider one specific result of this work only, which concerns
"anomalous"   momentum transfer between two quantum objects (i.e.~a photon and a mirror) and thus appears to
be in intimate connection with 
the neutron-atom collision (or scattering) experiments considered in the following sections.

\begin{figure}[ht]
\begin{center}
\includegraphics[width=120 mm]{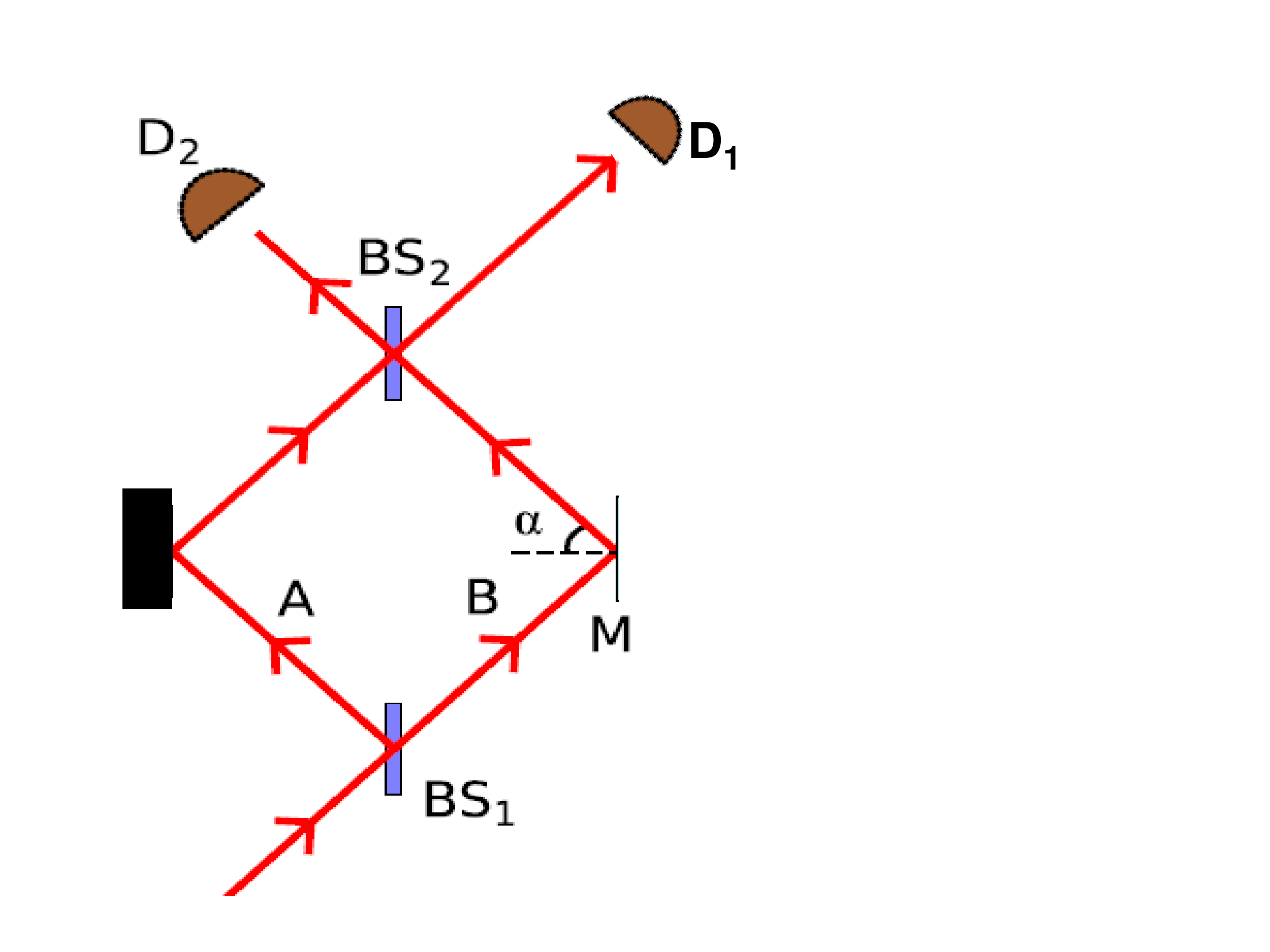}
\end{center}
\caption{{\label{Fig-NJPnew} Mach-Zehnder interferometer and path of a light beam (or a particle, etc.). 
The mirror M  
in the rhs is a  mesoscopic \textit{quantum}  object. Of particular interest are the photons (particles) 
emerging toward detector D$_2$. (Adapted from Fig.~2 of \cite{Aharonov-NJP}.)}  }  
\end{figure}

The schematic representation of the experimental setup is shown in Fig.~\ref{Fig-NJPnew}.
   
A photon (or particle) beam enters a device similar to a usual Mach-Zehnder interferometer (MZI), with the
exception that one reflecting mirror is sufficiently small (say, a meso- or nanoscopic object $M$) in order
that its momentum distribution may be detectable by a suitable non-demolition measurement
\cite{Scully1997}.   
  In  Ref.~\cite{Aharonov-NJP} the authors show  the following astonishing result.
Although the post-selected photons (as all photons do, of course) collide with the mirror $M$ only
from the \emph{inside} of the MZI, they do not push $M$ outwards, but rather they somehow succeed
to \emph{pull it in}.  It is obvious that this result cannot have any conventional theoretical
interpretation.            
 %As Aharonov et al.~put it: "This is realized by a superposition of giving the
 % mirror zero momentum and positive momentum
 %--- the superposition results in the mirror gaining negative momentum."

Let us now consider a straightforward derivation of this effect,  following the presentation 
of  some  derivations (relevant to our purposes)  of   Ref.~\cite{Aharonov-NJP} closely.  
   
 The two identical beam splitters of MZI
 have nonequal reflectivity $r$ and transmissivity $t$ (both real, with $r^2 + t^2 = 1$),
 say $r > t$.  Now we are interested in the momentum kicks given to the mirror $M$
 due to the photon-$M$ collisions inside the interferometer, but especially for
photons emerging toward detector D$_2$ in Fig.~\ref{Fig-NJPnew}).
 This corresponds to a post-selection condition.   (The effect of the photons emerging toward
 the other detector, D$_1$,  is of less relevance here.)

 Using the standard convention, an incoming
state $|{\rm in}\ra$ impinging on the beam splitter will emerge as a superposition of a reflected state $|R\ra$ 
and a transmitted state
$|T\ra$; $|{\rm in}\ra\rightarrow ir|R\ra+t|T\ra$. Hence when a single photon impinges from the left on BS$_1$,
as illustrated in Fig.~1, the effect of the beam splitter is to produce inside the interferometer the state
\beq|\Psi\ra=ir|A\ra+t|B\ra\eeq where $|A\ra$ and $|B\ra$ denote the photon propagating along the $A$ and $B$
arms of the interferometer, respectively.

The second beamsplitter, BS$_2$, is identical to the first. As one can readily check, a photon
in the quantum state $|\Phi_1\ra=t|A\ra-ir|B\ra $ impinging on the second beamsplitter, emerges towards
detector D$_1$ while a photon in the orthogonal state $|\Phi_2\ra=-ir|A\ra+t|B\ra$ emerges towards detector D$_2$.

Thus, when a single photon (particle)  enters the interferometer by impinging on the left side of the
beamsplitter BS$_1$, the probabilities to be found in the arms $A$ and $B$ are $r^2$ and $t^2$ respectively. The
probability of emerging towards detector D$_1$ is $\big |\la \Phi_1|\Psi\ra\big |^2=4r^2t^2$ while the
probability of emerging towards D$_2$ is $\big |\la \Phi_2|\Psi\ra\big |^2=(r^2-t^2)^2=1-4r^2t^2$.

Firstly, we send a \textit{classical} light (particle) beam of intensity $I$ towards this interferometer.  
The light (particle) intensity in arm $A$ is then   $I_A=r^2I$ while the intensity in arm $B$ is $I_B=t^2I$. 
The
intensities of the output beams, toward the tow detectors, are $I_{D_1}=4r^2t^2I$ and $I_{D_2}=(1-4r^2t^2)I$.
The momentum given to the mirror $M$ by the beam inside the interferometer is
$2I_B\cos\alpha=2t^2I\cos\alpha$. Clearly, this  pushes the mirror $M$  outward.

Secondly, let us now send a \textit{quantum} beam of photons (or particles) into the MZI. 
 Each photon
incident on $M$ gives it a momentum kick $\delta$. Note  that each individual momentum kick must be much smaller than the
spread $\Delta$ of momentum of the mirror $M$. This is a general property of any interferometer. It has to be 
so in order to maintain the coherence of the beam in
the interferometer; otherwise the photons (particles)  will become entangled with the mirror $M$. 
To show this, we note that a photon when going
through arm A will produce no kick to M while when going through arm B it will. 
 Accordingly, if $\phi(p)$ is the 
initial quantum state of $M$  and by $|\Psi\ra$ the quantum state of the photon after the input 
beam splitter BS$_1$, but before reaching the mirror, the reflection on the mirror results in
  \beq |\Psi\ra\phi(p)=(ir|A\ra+t|B\ra)\phi(p)\rightarrow ir|A\ra\phi(p)+t|B\ra\phi(p-\delta)\label{state}
	\eeq

If $\phi(p)$ is orthogonal to $\phi(p-\delta)$ where
$\delta$ is the kick given by the photon, then the photon ends up entangled with the
mirror and coherence is lost. Another way of looking at this is to note that the mirror has to be
localized within a distance smaller than the wavelength of light, otherwise there will be phase fluctuations
larger than $2\pi$ and interference is lost. 

[In fact the spread $\Delta$ in the momentum of the mirror has to be many times
bigger --- of order $\sqrt {\overline n}$ times - than that of an individual kick to ensure coherence when a beam
with an average of $\overline n$ photons and a spread $\sqrt {\overline n}$ goes through the interferometer. At
the same time $\Delta$, being of order $\sqrt {\overline n}\delta$, is small enough so that the average kick, which is of order
${\overline n}\delta$ is detectable.]
  %%%%%%%%%%%%%%%%%%%%%%%%%%%%%%%%%%%%%%%%%%%%%%%%%%%

For simplicity we take the state of the mirror to be (up to normalisation)
$\phi(p)=\exp({-{{p^2}\over{2\Delta^2}}})$.  Consider now a single photon propagating through the
interferometer. Given that $\delta\ll\Delta$, we can approximate the state (\ref{state}) of the photon and
mirror just before the photon reaches the output beamsplitter by 
 \barr &&|\Psi\ra\phi(p)
\approx
ir|A\ra\phi(p)+t|B\ra\Big(\phi(p)-{{d\phi(p)}
\over{dp}}\delta\Big)\nonumber\\&&=|\Psi\ra\phi(p)-t|B\ra{{d\phi(p)} \over{dp}}\delta  
\earr

Suppose now that the photon emerges in the beam directed towards detector D$_2$. 
The state of the mirror $M$ is then given    (up to normalization) by projecting the joint state 
onto the state of the photon corresponding to this beam,  i.e.
 \barr
&&\la\Phi_2|\Big(|\Psi\ra\phi(p)-t|B\ra{{d\phi(p)} \over{dp}}\delta\Big)\nonumber\\
&=&\la
\Phi_2|\Psi\ra\Big(\phi(p)-{{t\la\Phi_2|B\ra}\over{\la \Phi_2|\Psi\ra}}{{d\phi(p)}
\over{dp}}\delta\Big)\nonumber\\
&=&\la \Phi_2|\Psi\ra\Big(\phi(p)-{{\la\Phi_2|\textbf{P}_B|\Psi\ra}\over{\la
       \Phi_2|\Psi\ra}}{{d\phi(p)} \over{dp}}\delta\Big)\nonumber\\
&=&\la  \Phi_2|\Psi\ra\phi(p-P_B^w\delta)  
\label{mirrorstate}
 \earr 
Here $\textbf{P}_B=|B\ra\la B|$ is the \textit{projection operator} on
state $|B\ra$ and 
 $$P_B^w={{\la\Phi_2|\textbf{P}_B|\Psi\ra}\over{\la \Phi_2|\Psi\ra}}$$ 
is the so called \textit{weak value}  (WV) of
$\textbf{P}_B$ between the initial state $|\Psi\ra$ and the final state $|\Phi_2\ra$ 
\cite{Aharonov-Buch,AAV1988,Aharonov1990,Kofman2012,Tamir2013,Aharonov2014,Dressel2014}.
The value of $P_B^w$ is
readily found to be
    \barr P_B^w&=&{{\la \Phi_2| \textbf{P}_B|\Psi\ra}\over{\la \Phi_2|\Psi\ra}}={{(ir\la A|+t
\la B|)\textbf{P}_B(ir|A\ra+t |B\ra)}\over{(ir\la A|+t \la B|)(ir|A\ra+t
|B\ra)}}\nonumber\\&=&-{{t^2}\over{r^2-t^2}}  
  \earr 
	Hence the momentum kick received by the mirror due to a
photon emerging towards D$_2$ is 
\beq\delta p_M=P_B^w\, \delta=-{{t^2}\over{r^2-t^2}}\, \delta  
\eeq

The appearance in the above expressions of the WV of the projector $\textbf{P}_B$ is
not accidental. Indeed, we can view the mirror as a measuring device measuring whether or not the photon is in
arm B or not. The momentum of the mirror acts as a ``pointer'' (no kick --- the photon is in arm A; kick --- the
photon is in arm B). However, since the photon can only change the position of the pointer (i.e. the momentum
of the mirror) by far less than its spread, we are in the so called "weak measurement"
\cite{Aharonov-Buch,AAV1988,Aharonov1990,Kofman2012,Tamir2013,Aharonov2014,Dressel2014}    regime.

Finally, the total momentum given to the mirror by all the photons emerging towards D$_2$ is given by the
momentum due to each photon times the number of photons in the beam. Using the fact that the probability of a
photon to end in this beam is $(r^2-t^2)^2$ we obtain
\beq \delta p_M={\overline n}(r^2-t^2)^2{{-t^2}\over{r^2-t^2}} \delta =-t^2
{\overline n}(r^2-t^2)\, \delta < 0 
\eeq 
Since we investigated the case with
$r>t$, the sign of the momentum received by the mirror is negative, hence the mirror is pushed towards the
inside of the MZI. This momentum change is a result of the mirror receiving a {\it
superposition} between a kick $\delta$ and no kick at all, corresponding to the photon (particle) 
propagating through the two MZI-arms.

%%%%%%%%%%%%%%%%%%
To summarize: The physical insight obtained from quantum mechanics is dramatically 
different from the classical one. 
In fact, according to quantum mechanics, the photons that end up in the D$_2$
beam  give  negative momentum kick on $M$. Astonishingly, although they collide with the mirror only from the inside of 
the MZI, they do not push the mirror outwards; rather they somehow succeed to pull it in! This is
realized by a superposition of giving the mirror zero momentum and positive momentum --- the superposition results
in the mirror gaining negative momentum.

%%%%%%%%%%%%%
Concluding the above short discussions, one may say that the new insights and/or predictions made
possible within the theoretical frame of WV and TSVF are not limited to interpretational issues
only. The revised intuitions can lead one to find  novel quantum effects that can be measured
in real experiments.

\section{On post-selection, weak measurement, weak values and two-state-vector   formalism }

   %%absract Aharonov2014
Weak quantum measurement (WM) is unique in measuring noncommuting operators and
revealing new counter-intuitive
effects predicted by the two-state-vector-formalism (TSVF)
 \cite{Aharonov1990,Aharonov2014}.
The main aim of this article is 
to point out certain  new (and experimentally observable) features of
elementary scattering processes
predicted within the theoretical frame of  WM and TSVF, and which contradict every conventional
 expectation; cf.~\cite{Dreismann-Quanta}. Concretely, we have in mind
incoherent-inelastic scattering of single (massive) particles,
(e.g. neutrons or electrons) from nuclei and/or atoms. 

As the starting point of the new theory under consideration one usually considers
the paper \cite{AAV1988} by Aharonov, Albert and Vaidman, and
the earlier paper \cite{ABL1964} by Aharonov, Bergmann and Lebowitz.
Here some short remarks may be helpful.

A standard von Neumann (also called "strong") measurement yields the eigenvalues of the measured
observable, but at the same time  disturbs the
measured system. According to standard theory  \cite{vonNeumann},
 the final state of the
system after the measurement becomes an eigenstate of the measured
observable. This usually changes the initial state of the system.
On the other hand, by coupling a measuring device to a system sufficiently weakly,  it
 may be possible to read out certain information while limiting the
disturbance induced by the measurement to the system. As Aharonov and collaborators
originally proposed  \cite{AAV1988,ABL1964},
 one may achieve new  physical insights  when one furthermore
 post-selects on a particular outcome
 of the experiment. In this case the eigenvalues of the measured observable
are no longer the relevant quantities; rather the measuring device
consistently indicates the {\em weak value} (WV) \cite{AAV1988,ABL1964} given by
   \be \label{AAV-formula}
A_w \equiv  \frac{\bra{\psi_f} \hat{A} \ket{\psi_i}}{\braket{\psi_f}{\psi_i}}
  \ee
where $\hat{A}$ is the operator whose value is being ascertained,
$\ket{\psi_i}$ is the initial state of the system, and $\ket{\psi_f}$
is the state that is post-selected (e.g. by performing a specific
measurement). Note that the number $A_w$ may be complex.

The significance of this formula is as follows.
Let us couple a
measuring device whose pointer has position coordinate $q$ to the
system $\cS$, and subsequently measure its conjugated momentum $p$.
The coupling interaction is
taken to be the standard von Neumann measurement interaction \cite{vonNeumann}
 \be
\hat{H}=-g \, \hat{q} \otimes \hat{A}
 \label{interaction}
 \ee
 The coupling constant $g$ is assumed to be appropriately small and the
 interaction time sufficiently short.
Then the mean value
$\langle p \rangle$
of the pointer momentum is given by \cite{AAV1988}
\be \label{p-average} \langle p \rangle = g\ Re [A_w]
\label{ReWV}
 \ee
where $Re$ denotes the real part. This formula requires the initial
pointer momentum wavefunction to be real and centered at $\langle p \rangle =0$
before measurement, but these
assumptions can easily be relaxed; see cited references.

The formula (\ref{AAV-formula}) implies that, if the initial state
$\ket{\psi_i}$ is an eigenstate of a measurement operator $A$, then
the weak value post-conditioned on that eigenstate is the same as the
classical (strong) measurement result. When there is a definite
outcome, therefore, strong and weak measurements agree. Interestingly, a
WM  can yield values outside the  range of measurement
results predicted by conventional theory \cite{AAV1988}.

For some experiments,  but not in those considered in this paper, 
a WV can also be complex, with its
 imaginary part affecting the pointer position. I.e.,
the mean of the pointer position after measurement is given by
\be \label{q-average}
\langle q \rangle = 2gv\ Im [A_w]  
\ee
where $Im$ denotes the imaginary part and $v$ is the variance in
the initial pointer spatial position \cite{AAV1988}, with
$\langle q \rangle=0$ before measurement.

The situation we shall consider is where a system $\cS$ evolves
unitarily from an initial state $\ket{\psi_i}$ to a final
post-selected measurement outcome $\bra{\psi_f}$. At various time points
inbetween,
observables may be measured weakly. Here we consider the scenario
where there is a single copy of the system, with the measuring device
weakly coupled to it.

In the simplest case where there is just one observable $A$, we
assume the evolution from $\ket{\psi_i}$ to the point where $A$ is
measured is given by $U$, and from this point to the post-selection
the evolution is given by $V$. Then we can rewrite (\ref{AAV-formula})
as:
\be \label{fullAAV}
A_w=\frac{\bra{\psi_f}\hat{V}\hat{A}\hat{U}\ket{\psi_i}}{\bra{\psi_f}\hat{V}\hat{U}\ket{\psi_i}} 
\ee
and the mean of the pointer is given by (\ref{p-average}) as before.

The fact that one only sufficiently "weakly" disturbs the system in making WMs
 means that one can in principle measure different
variables in succession. This theoretical observation has led to a great number of experimental
applications and discovery of several new effects; see e.g.   
 \cite{Aharonov-Buch,AAV1988,Aharonov1990,Kofman2012,Tamir2013,Aharonov2014,Dressel2014,Dreismann-Quanta}
and references cited therein.

%%%%%%%%%%%%%%%%%%%%%%%%%%%%%%%%%%%% %%%%%%%%%%%%%
\section{\label{sec:4} Elementary scattering in view of WV and TSVF}

\subsection{\label{sec:4.1} Momentum transfer in impulsive two-body collisions}

In this theoretical section, we present the basic result of WV and TSVF as applied to scattering
processes, especially non-relativistic neutron scattering, 
firstly obtained in Ref.~\cite{Dreismann-Quanta}. 

Here we mainly follow the presentation of that reference. 

The
position and momentum of the neutron (probe particle) are denoted as $(q,p) $. Similarly, the
position and  momentum of the scatterer (atom, nucleus) are denoted as $(Q,P)$. 
$\hat{X}$ may represent the corresponding operator of a physical quantity $X$. For simplicity,
 let us consider here a  one-dimensional quantum model for
momentum exchange in a two-particle impulsive collision.

In the usually considered, but extremely limiting case in which both particles occupy 
states with 
well defined momenta (i.e. plane waves),
the initial state of the two-body system is standardly assumed as a product (uncorrelated) state
\begin{equation}
\Psi_{in}
=
\phi_{n}(p) \otimes  \Xi_{A}(P)
\label{initial state}
\end{equation}
(indices $n$ and $A$ refer to neutron and atom, respectively).
In a first attempt, an impulsive scattering process  may be formally approached by 
the  interaction Hamiltonian
\begin{equation}
\hat{V}     =  F(t) \, ( \hat{q} - \hat{Q} ) 
\end{equation}
where the function $ F(t) $ represents  a non-zero force  during a
short time interval $\tau$, the duration of the collision.
We may assume  that $ F(t) $ is proportional to a delta
function. 
Furthermore,  the integral  
\begin{equation}
\int_0^\tau F(t) \, dt =  \hbar K
\end{equation}
provides the  momentum transfer $ \hbar K$ caused by the collision.

Neutron and atom observables commute, so
$ [ \hat{q} , \hat{Q} ] =0$, and the associated
unitary evolution operator is
\begin{equation}
\hat{U}(\tau) = e^{ - (\imath/\hbar) \int \hat{V} dt }
=
e^{ - (\imath/\hbar)\, \hbar K \,  (\hat{q} - \hat{Q}) }
\equiv
e^{- \imath \, K \, \hat{q}} e^{+ \imath \, K \, \hat{Q}}  
\end{equation}
The \textit{shift operator} $ e^{ \imath \hbar K \,\hat{Q}/\hbar } $ 
describes the momentum shift of an atomic momentum eigenket as
$e^{\imath \hbar K \, \hat{Q} / \hbar} | P \rangle = | P + \hbar K \rangle $
while the operator
$e^{-\imath \hbar K \, \hat{q} / \hbar} $ shifts neutron's
momentum eigenket
as $ e^{- \imath \hbar K \hat{q} / \hbar} | p \rangle = | p - \hbar K \rangle $  
of the impinging particle (neutron).
Immediately after the momentum exchange, the state of the two-particle system in the momentum
representation is
\begin{eqnarray}
\Psi_f &=&
\hat{U}(\tau) \, \phi_{n}(p) \otimes \Xi_{A}(P)
\nonumber \\
&=& e^{-i \hbar K \, \hat{q} / \hbar}  \phi_{n}(p)
\otimes
e^{ i \hbar K \, \hat{Q} / \hbar}  \Xi_{A} (P) \nonumber \\
&=& \phi_{n} (p + \hbar K) \otimes   \Xi_{A}(P - \hbar K)  
\label{conventional}
\end{eqnarray}
which is a not entangled final state,  due to the trivial form of
$\hat{V}$.
These  introductory considerations  may motivate the search for a more
appropiate (i.e. more realistic) two-body
impulsive interaction Hamiltonian, which is presented in the
following section.

\subsection{\label{intHam} WM, weak interaction Hamiltonian and momentum transfer}

Let us now assume that the struck atom is initially at rest. 
For simplicity of notation, it is sufficient to
consider the atomic momentum component parallel to $\textbf{K}$, which from now on we shall denote
simply by $P$---and the associated operator simply by $ \hat{P}$. 
In other words, we consider a one-dimensional problem.

For illustration and motivation of the derivations presented below, we  first provide 
a heuristic
derivation of a von Neumann-type \cite{vonNeumann}  interaction 
Hamiltonian for  momentum transfer.
Let us start with a formal one-body model  Hamiltonian describing momentum transfer $-\hbar K\equiv +\hbar
K_n$   \emph{on the neutron} due to the collision with the atom:
\begin{equation}
\hat{V}_n(t)= \delta(t)\, \hbar K \, \hat{q}  
\end{equation}
The associated evolution operator acting on the space of the neutron
 \begin{equation}
\hat{U}_n(\tau) =
\exp ( - \frac{\imath}{\hbar}\, \hbar K \,  \hat{q} )
\label{Un}
\end{equation}
shifts a   momentum eigenstate of the impinging particle (neutron) as $ e^{- \imath \hbar K
\hat{q} / \hbar} | p \rangle = | p - \hbar K \rangle $.

Due to momentum conservation in the two-body collision, one has
\begin{equation}
- \hbar  K_n = \hbar K_A \equiv  \hbar K  
\label{K-definitions}
\end{equation}
where $\hbar K_A$ is the momentum transfer  \emph{on the atom} due to the collision.
(We choose $K_A$ as positive, following standard notation; 
see e.g.~the textbook \cite{Squires} or the review article \cite{Watson}.)
   
Let the scattering atom be at rest before collision, $\langle
\hat{P} \rangle_{i} = 0 $. Hence, after the collision,  
    it holds
\begin{equation}
\hbar K_A =+\hbar K = \langle \hat{P} \rangle_{f} =
\langle \hat{P}\rangle_{f} - \langle \hat{P}\rangle_{i}
\label{Pconventional}
\end{equation}
%%%%%%%%%%neu Aug
and correspondingly for the neutron momentum
\begin{equation}
\hbar K_n =-\hbar K =
\langle \hat{p}\rangle_{f} - \langle \hat{p}\rangle_{i}  
\label{Pconventional2}
\end{equation}
%%%%%%%%%%%%%%%%%
Hence, the aforementioned operator $\hat{U}_n(\tau)$ of the neutron, Eq.~(\ref{Un}),
may be written as
\begin{equation}
\hat{U}_n(\tau) =
\exp(  - \frac{\imath}{\hbar}\, \langle \hat{P}\rangle_{f}
\,  \hat{q} ) 
\label{U2}
\end{equation}

To apply the theory of weak measurement (WM) and two-state-vector formalism (TSVF), 
a  \textit{von Neumann} two-body
interaction Hamiltonian is needed; see the associated references cited in the Introduction. 
Thus one is  guided to search of a two-body generalization of the
one-body evolution operator $\hat{U}_n(\tau)$ of the form
\begin{equation}
\hat{U}(\tau) =
\exp( - \frac{\imath}{\hbar}\,  \hat{q} \, \hat{P} ) 
\label{U3}
\end{equation}
However, this heuristically derived expression  still has  no  direct context to the \emph{realistic}
experimental situations under consideration (see Sec.~6 below).  
To  achieve this,  we now may proceed as follows.

To begin with, let us refer to the so-called 
\textit{impulse approximation} (IA) of standard theory \cite{Watson,Hohenberg,Sears} and Eq.~(\ref{E-conserv})
regarding energy conservation  (see Sec.~5):
$$
E=  \frac {(\hbar K)^2}{2M} +\frac{\hbar{\mathbf K\cdot \mathbf P}}{M}  
$$
Looking at this equation, and having the WV and TSVF in mind,  
one sees that the \emph{larger} recoil term $\frac {(\hbar K)^2}{2M}$ may be
viewed to result from a \emph{strong} impulsive interaction (associated with momentum transfer $+\hbar
K$ on the atom). The theoretical treatment of this part of the interaction can be found in standard textbooks
(e.g. \cite{Squires}) but is   not 
within  the scope of the present paper. 
Since in the impulse approximation (IA) holds $|K| \gg |P|$,
the \emph{smaller} Doppler term $\frac{\hbar{\mathbf K\cdot \mathbf P}}{M}$ may correspond to a
\emph{weaker} interaction, in which the atomic momentum
$\hat{P}$
couples with  an appropriate dynamical variable of the neutron, say ${\cal{\hat{O}}}_n $.   
Looking  at the preceding formulas
(\ref{U2},\ref{U3}) for the model operator effectuating momentum transfer, it appears that this
dynamical variable should be  $\hat{q}$, that is ${\cal{\hat{O}}}_n  = \hat{q}$.   

In view of the theory of WV and TSVF, the  weak
interaction is expected to cause weak \emph{deviations} from  the conventionally 
expected \emph{large} momentum transfer $\hbar K$. This can be
formally captured by replacing $ \hat{P}$ with the \emph{small} momentum difference
$\hat{P} - \hbar K \hat{I}_A  $, and also including a positive \emph{smallness} factor
$$0<\lambda \ll 1$$
in the model interaction Hamiltonian. Summarizing these considerations, let us
assume the model interaction Hamiltonian
\begin{equation}
\hat{H}_{int}(t) = + \lambda \,\delta (t) \,  \hat{q}
\otimes  (\hat{P} - \hbar K \,  \hat{I}_A )   
\label{H-int}
\end{equation}

It should be pointed out that the \emph{plus sign} in front of this expression is \textit{not} arbitrary,
since it is consistent with the aforementioned definitions (\ref{K-definitions}) of momentum transfer.
This point    plays a decisive role in the context of
the new quantum effect of momentum transfer deficit.

For further physical motivation of the two parts of the model Hamiltonian of Eq.~(\ref{H-int}), it
may be helpful     to compare the above reasoning with an example by Aharonov et
al.~\cite{AharonovEPJ2014}, p.~3: 

"Consider, for example, an ensemble of electrons hitting a nucleus in a particle
collider. $\left[\ldots\right]$
Their initial states are known, and a specific post-selection is done after the interaction.
The main interaction is purely electromagnetic, but there is also a relativistic and
spin-orbit correction in higher orders which can be manifested now in the form of a weak
interaction."

\subsection{\label{deriv} WV of atomic momentum operator $\hat{P}$ and the 
effect of "anomalous" momentum transfer}

Here, the atom represents the system of the general formalism.
Since the WV of the identity operator is
$ (\hat{I}_A )_w = 1$,
for the WV of the  atomic coupling
operator  $\hat{P} - \hbar K\,\hat{I}_A $ in the above interaction Hamiltonian holds:
\begin{equation}
(\hat{P} - \hbar K \,  \hat{I}_A )_w = P_w - \hbar K   
\end{equation}
To proceed,  we first calculate the WV $P_w$ of $\hat{P}$ for some characteristic (and
experimentally relevant)  final state in momentum space.
The derivation is rather straightforward and reveals a striking
deviation --- more precisely, a deficit --- from the conventionally 
expected momentum transfer to the neutron. The latter represents the pointer of general theory,
and  the pointer momentum variable $\hat{p}$ is conjugated to $\hat{q}$ contained in 
Eq.~(\ref{H-int}).

Moreover, for the calculation of the WV, it is natural to use the momentum
space representation, as scattering experiments usually measure
momenta (rather than the positions of the scatterers in real space).

Let the atom initially be in a spatially confined state and at rest; 
e.g. in a potential representing physisorption on a surface;
cf.~examples  in experimental section below. Usually, the initial atomic wave
function $\Xi(P)_i$ is approximated by  a Gaussian $G_A$
centered at zero momentum,
$$\Xi(P)_i \approx  G_A(P) .$$
At sufficiently deep temperature the atom will be in its ground state, and the width of $\Xi(P)_i$
is determined by the quantum uncertainty.

It follows that the struck atom moves in the direction of momentum transfer $\hbar K_A = \hbar K$; therefore, to
simplify notations, in the following calculations $P$ represents the  atomic momentum along the
momentum transfer direction.

To be specific, as well as to facilitate the derivations, let us make the following simplifying 
assumption concerning the final atomic state: 
\begin{itemize}
\item{The final atomic state has the same width in momentum space as the initial state.}
\end{itemize}
However, it may be noted that this assumption is very common in molecular (optical) spectroscopy.
It captures the viewpoint that the impulsive transition is very fast and so the atomic 
environment didn't have sufficient time to change configuation and adapt to the "disturbance" due to
the atomic final state. 

In other words,   let the final atomic state have the same shape as the initial
state, but its center should be shifted from zero to  the transferred momentum, i.e. 
\begin{equation}
\Xi(P)_f =  \Xi(P-\hbar K_A)_i .
\end{equation}
The WV of the atomic momentum operator $\hat{P}$ is calculated as follows:
\begin{eqnarray}
P_w &=& \frac{\bra{\Xi_f} \hat{P} \ket{\Xi_i}}{\braket{\Xi_f}{\Xi_i}}
\nonumber  \\
&=& \frac{\int dP\, \Xi(P-\hbar K_A)_i \, P    
\Xi(P)_i}
{\int dP\,  \Xi(P-\hbar K_A)_i\, \Xi(P)_i}
\nonumber  \\
&=& + \frac{\hbar K_A}{2} \nonumber \\
&=& + \frac{\hbar K}{2}   
\label{deltaK-C}
\end{eqnarray}
The last equality  follows immediately from $(a)$ the two symmetrically distributed $\Xi$
functions around their central position
$\bar{P} =\hbar K_A/2$
and $(b)$ the linear term  $P$ in the integral of the
nominator. 

It should be noted that this result does not depend on the
width of $\Xi$, as long as the two $\Xi$ functions are not orthogonal to each other.

Thus we arrived at the following striking conclusion: 
The momentum transfer comes out to be only half of the conventionally expected 
value $\hbar K$. 

This is a quite interesting result because it represents a
momentum-transfer deficit of 50\%; i.e. the scattered neutron
measures a momentum kick being only half of the conventionally
expected value. In more detail, it holds
\be
(\hat{P} - \hbar K \,  \hat{I}_A )_w =  +\frac{\hbar K_A}{2} -\hbar K = -\frac{\hbar K}{2}
\ee
To proceed, one should   note the change of sign between the usually applied interaction
Hamiltonian of Eq.~(\ref{interaction}), as e.g.~treated in \cite{AAV1988}, and the
presently used specific form (\ref{H-int}). Therefore, there is a corresponding change of sign in
 Eq.~(\ref{ReWV}), 
\begin{equation}
\langle \hat{p} \rangle_f - \langle \hat{p} \rangle_i=  -  g\, \textrm{Re} [A_w]  
\label{changed-sign}
\end{equation}
Thus   one obtains for the \textit{correction}   
to the  shift of the meter pointer variable:
\begin{equation}
\langle \hat{p}\rangle_f -  \langle \hat{p}\rangle_i = -\lambda\, (\hat{P} - \hbar K \,  \hat{I}_A )_w
= + \lambda\,\frac{\hbar K}{2}  
\label{half}
\end{equation}
    Moreover,
    for the
\emph{total} momentum transfer shown by the pointer of the measuring device (here: of neutrons) we have
\begin{eqnarray}
\left[ \langle \hat{p}\rangle_f -  \langle \hat{p}\rangle_i \right]_\textrm{total} \ \
&=& \left[ \langle \hat{p}\rangle_f -  \langle \hat{p}\rangle_i \right]_\textrm{conventional} \nonumber \\
& &  \ \ \ \ \ \ \  + \left[ \langle \hat{p}\rangle_f -  \langle \hat{p}\rangle_i \right]_\textrm{correction} \nonumber \\
&=&  - \hbar K    + \lambda\,\frac{\hbar K_A}{2} \ \
\label{WM-result}
\end{eqnarray}
This expression represents the
new quantum effect of
\emph{momentum-transfer deficit}:  the absolute value of  momentum transfer
on the neutron
predicted by the new theory is \textit{smaller}
than that predicted by  conventional theory: 
\be
| - \hbar K  + \lambda\,\hbar K_A/2 | \leq | - \hbar K | 
\ee
Recall that by definition holds $\hbar K_A = \hbar K$; see Eq.~(\ref{K-definitions}).
 %%%%%%%%%%%%%%%%%%%%%%%%%%%%%%%%%%%%%%%%%%%%%%%%%%%%%%%55 RECHNUNG...

\subsubsection{Plane waves --- The limiting case of conventional momentum transfer.}

Moreover, it is interesting to point out  the succinct reason of this large "anomaly" by the following
calculation. Let us now make the usual assumption of  conventional theory 
(see e.g.~\cite{Squires,Watson,Hohenberg}) which is:
\begin{itemize}
 \item{The final state should be a plane wave (i.e. it has vanishing width
in momentum space) --- as assumed in general conventional theory and the IA.}
\end{itemize}
Here we will show straightforwardly that the result of conventional theory is reproduced.

Namely, now the final state is a delta function (in momentum space), 
that is,  
 the momentum wave function is a
delta function $\delta_A$  centered at the assumed transferred
momentum $\hbar K_A$,
\begin{equation}
\Xi(P)_f =  \delta_A(P-\hbar K_A) .
\end{equation}
The weak value of $\hat{P}$ follows straightforward:
\begin{eqnarray}
P_w &=& \frac{\bra{\Xi_f} \hat{P} \ket{\Xi_i}}{\braket{\Xi_f}{\Xi_i}}
\nonumber  \\
&=& \frac{\int dP\, \delta_A(P-\hbar K_A) \, P  
\Xi(P)_i}
{\int dP\,  \delta_A(P-\hbar K_A)\, \Xi(P)_i}
\nonumber  \\
&=&  \frac{\hbar K_A \, \Xi(\hbar K_A)_i }
{\Xi(\hbar K_A)_i}
\nonumber   \\
&=& + \hbar K_A \equiv + \hbar K 
\label{Pw-IA}
\end{eqnarray}
(Recall the notations of Eq.~(\ref{K-definitions}).) Hence, the WV of the system coupling operator
$(\hat{P} - \hbar K \, \hat{I}_A )$ is just zero,
\begin{equation}
(\hat{P} - \hbar K \,  \hat{I}_A )_w = P_w - \hbar K =0 
\label{noanomaly}
\end{equation}

[According to standard quantum scattering theory, the scattered wave
may acquire an additional phase factor, say $e^{\imath \chi}$, which does not affect the
preceding result because this factor cancels out in the fractions of Eqs.~(\ref{Pw-IA}).]

This physically means that, under the usual assumption of "plane waves", 
the new theory yields \textit{no correction} to the
conventionally expected value of momentum transfer:
\begin{equation}
\langle \hat{p}\rangle_f  -\langle \hat{p}\rangle_i = -\lambda \,(\hat{P} - \hbar K \,  \hat{I}_A )_w
= 0   
\label{no-deficit}
\end{equation}
 And  hence the
pointer of the measuring device will show the conventionally expected value $- \hbar K $.
Thus, the result of Eq.~(\ref{noanomaly}) is consistent with
conventional theory of the impulse approximation (IA, see below)
 and, more generally, of incoherent neutron
scattering.   
Clearly, as the standard result of conventional theory is now reproduced, 
this is a very satisfactory result.

Comparison of the aforementioned two derivations shows immediately that the degree of the momentum
transfer "anomaly" under considerations depends on the "deformation" of the shape of the
final atomic state. E.g., if the final state is, say, "nearly" a delta function (plane wave),
then the anomalous momentum transfer deficit will be "small" --- and perhaps remain undetectable.

The rather \emph{general} scheme of the above derivations provides evidence that the new
effect under consideration is  of a general character and thus  may apply to any field of scattering physics, 
(perhaps even to relativistic scattering in
high-energy physics). The immediate implication is that the general theory of WV and
TSVF may  be relevant for a broad range of modern scattering
experiments.

%%%%%%%%%%%%%%%%%%%%%%%%%%%%   EXPERIMENT   ##################--------------------------------------------

\section{Two-body collision --- Momentum- and energy-transfer}

In a typical scattering experiment, the involved 
amounts of energy and momentum transfer are determined. (Angular momentum or spin measurements 
play no role in this paper.)
However, this determination 
involves not only directly measured quantities, but also several theoretically motivated
"beliefs" and/or insights based on some "commonly applied" theory. Let us have a short
look at these issues.

\begin{figure}[ht]
\begin{center}
\includegraphics[width=100 mm]{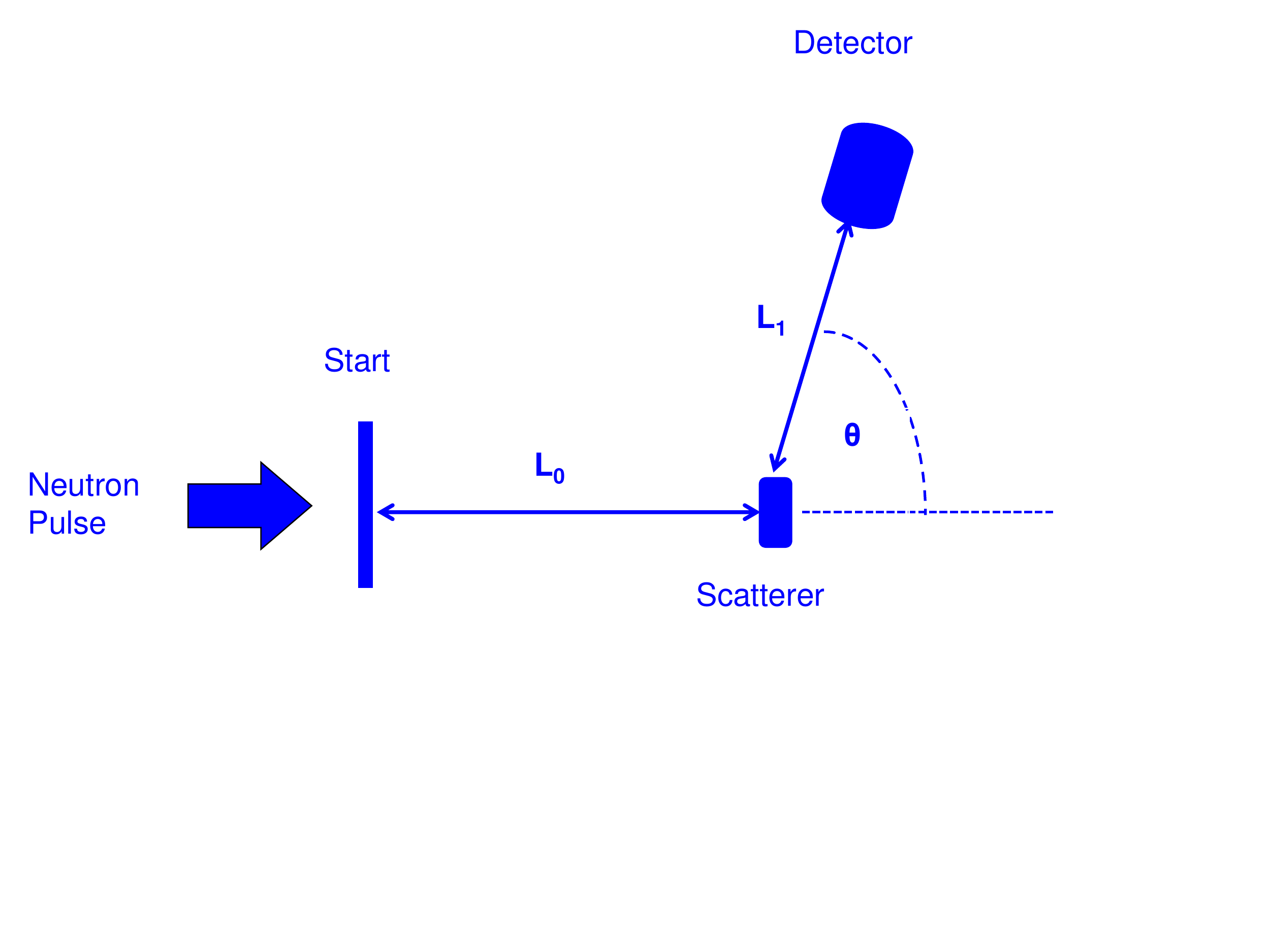}
\end{center}
\caption{{\label{TOF} Schematic presentation  of a time-of-flight scattering experiment.}}
\end{figure}

Modern neutron spectrometers
are time-of-flight (TOF) instruments; cf.~Fig.~(\ref{TOF}).  Here,
a short pulse of neutrons produced at a spallation source (e.g.~SNS or ISIS) 
reaches the first monitor of the spectrometer which triggers the measurement of TOF. 
A neutron scatters from the sample and may reache the detector, which stops the TOF measurement. 

Note that instrumental details play a crucial role in the
theoretical framework under consideration, because they concern pre-selection
and, in particular, post-selection being essential for the TSVF. 
Therefore the following facts should be pointed out.

$(a)$  From the measured TOF-values, but without using the actual value of scattering
angle $\theta$, one determines the value of $k_1 =  | {\mathbf k}_1|$,  and consequently of
energy transfer $E=\hbar \omega$; see Eq.~\ref{E-conserv}.

$(b)$ Momentum transfer $\hbar {\mathbf K}$, as defined in  Eq.~(\ref{K-vector}), is determined from both 
the scattering angle $\theta$ and the TOF value.

$(c)$ Furthermore, from each TOF-value measured with the
detector at $\theta$, the associated transfers of momentum ($\hbar K
$) and energy ($E=\hbar \omega$)   of the neutron to the struck
particle are uniquely determined; see e.g.~\cite{Dreismann-Quanta}. 
Hence a specific detector
measures only 
one specific trajectory in the whole $K$--$E$~plane.
Obviously, this fact is related  to the \emph{post-selection} of the WV and TSVF theory
under consideration

  %%%%%%%%%%%%%%%%%%%%%%   TOF   FIGURE   %%%%%%%%%%
In the following sections of this paper, the preceding theoretical considerations and results are
applied to concrete neutron scattering experiments, especially to \textit{incoherent} scattering. 
In simple terms, "incoherent" means that the impinging neutron (more generally: photon, electron, atom, etc.) 
collides with, and scatters from, a \textit{single} particle (nucleus, atom, molecule,  etc.) 
As concerns neutron scattering from protons (commonly referred to also as H-atoms), the 
scattering is mainly incoherent due to the spin-flip
mechanism of neutron-proton collision; see e.g.~\cite{Squires}.
   A clear first-principles explanation  of
\textit{coherent} versus \textit{incoherent} scattering 
may be found in Section 3-3 of the well-known Feynman Lectures  \cite{Feynman1965}.

Due to  scattering, the impinging neutron
causes a momentum transfer $\hbar \textbf{K} $ and an energy transfer $E$ to the atom. 
For simplicity, in the following
we consider isotropic scattering only, which is overwhelmingly dominant in non-relativistic 
neutron scattering \cite{Squires}. These quantities are
experimentally determined by standard methods; see 
 e.g.~Refs.~\cite{Dreismann-Quanta,Squires,Watson,Tietje,Sears}.

Due to  momentum conservation, it follows that
the neutron receives the opposite momentum $- \hbar \textbf{K} $.
The elastic collision of a neutron and a (free)  atom with mass $M$
and initial momentum $\textbf{P}$ results in the neutron's lost
energy $E\equiv\hbar \omega$ being transferred to the struck atom:
\begin{eqnarray}
E= E_i- E_f = \hbar \omega & = &
\frac {(\hbar \mathbf K + \mathbf P )^2}{2M} -\frac{P^2}{2M}  \nonumber \\
& = & \frac {(\hbar K)^2}{2M} +\frac{\hbar{\mathbf K\cdot \mathbf P}}{M} 
\label{E-conserv}
\end{eqnarray}
$E_i$ and $E_f$ are   the  neutron's initial and final kinetic energy, respectively.
 This equation represents energy conservation. Furthermore,
 \begin{equation}
\hbar {\mathbf K}  =\hbar {\mathbf k}_i - \hbar {\mathbf k}_f 
\label{K-vector}
\end{equation}
where   ${\mathbf k}_i$ and ${\mathbf k}_f$ are the   neutron's initial and final wavevectors, respectively.
The absolute value of  ${\mathbf K} $ is given by 
\begin{equation}
| {\mathbf K} |= K =\sqrt{k_i^2+k_f^2 - 2k_ik_f\cos \theta} 
\label{K}
\end{equation}

The first term in the right-hand-side (rhs) of Eq.~(\ref{E-conserv}) is the so-called recoil energy, 
\begin{equation}
E_{rec}=\hbar\omega_{rec} = \frac{(\hbar K)^2}{2M}   
\label{recoil}
\end{equation}
and represents the kinetic energy of a recoiling  atom being
initially at rest, i.e.~ $\langle P \rangle=0$. Hence one may write
\begin{equation}
\langle E \rangle = \frac{ \hbar^2 \langle K \rangle^2 }{2M} \equiv  \bar{E}_{rec} 
\label{E-conserv2}
\end{equation}
which holds for the $E$ and $\mathbf K$ of the  peak center.
Thus incoherent scattering from (say, a gaseous sample of) such atoms leads to a experimental 
recoil peak centered
at energy transfer  $\bar{E}_{rec}$, and exhibiting a
width being caused by the term $\hbar {\mathbf K \cdot \mathbf P}/M$,  which represents the well known effect of
\emph{Doppler broadening} --- being the second term in the rhs of Eq.~(\ref{E-conserv}).
  \begin{equation}
E_{Doppler} = \frac{\hbar{\mathbf K\cdot \mathbf P} }{ M} \equiv \frac{\hbar K P_\|}{ M} 
\label{Doppler}
  \end{equation}
where  $ P_\| $ denotes the atomic-momentum component  parallel to $\textbf{K}$. For
\emph{isotropic} systems (gases, liquids, amorphous solids etc., as those considered
below) the specific direction determined by
$\textbf{K}$ becomes immaterial, and thus $P_\| $ may represent the projection along 
any direction.

The above simple formulas of this section capture the features of the so-called \textit{Impulse
Approximation} (IA) \cite{Watson,Sears,Tietje,Hohenberg}.  

\begin{figure}[ht]
\begin{center}
\includegraphics[width=80 mm]{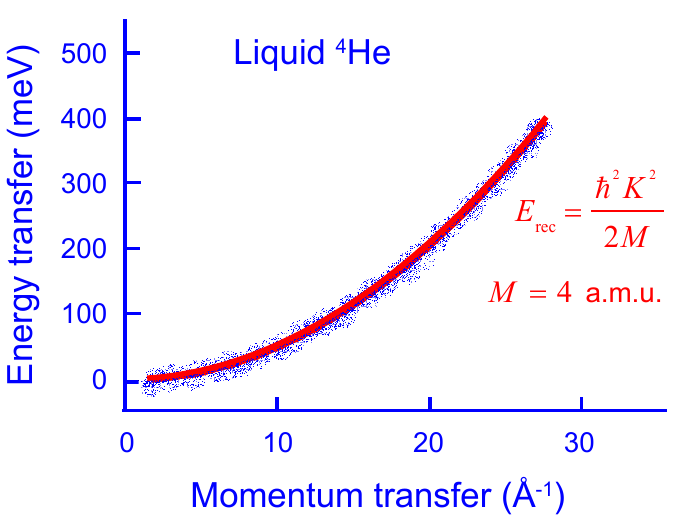}
\end{center}
\caption{\label{He-recoil} 
Schematic representation (blue points) of measured dynamic structure factor $S(K,E)$ of liquid helium \cite{Glyde}. 
The red line is the calculated recoil parabola,  according to   
 Eq.~(\ref{recoil}), for the mass of $^4$He, shown as a guide to
the eye. The white-blue ribbon  around the recoil parabola represent data points measured with
the time-of-flight spectrometer ARCS \cite{ARCS}  (Figure taken from Ref.~\cite{Dreismann-Quanta} with permission 
from Quanta.) }
\end{figure}

Relation (\ref{recoil}), also called  recoil parabola, is nicely illustrated in Fig.~\ref{He-recoil}, 
which shows experimental data of
incoherent inelastic (Compton, DINS) scattering from $^4$He atoms in  
the liquid phase; see \cite{Glyde} for details.

                          %%%%%%%%%%%%%%%%%%%%%%%%%%%%%%%%%%%%%%%%%%    
\subsection{\label{EffectiveMass}Effective mass as measured in the scattering experiment}

In real experiments, deviations from the impulse approximation are well known; 
see e.g. \cite{Watson,Tietje,Sears}. 
Within conventional theory such deviations are understood as follows.  

The energy conservation relation (\ref{E-conserv}) for a
two-body collision
holds in the so-called impulse approximation, which holds exactly for infinite momentum transfer 
and, consequently,  for quasi-free scattering particles.
But it is
not completely fulfilled at finite  momentum and energy transfers of actual experiments, and
thus so-called \emph{final-state effects} (FSE) may become apparent \cite{Watson,Tietje,Sears}. 
(In fact, this  term commonly includes both \emph{initial and final} state effects).   
FSE are caused by environmental interactions with the struck particle, which affect both initial and
final states of it. Here, we shall shortly discuss this effect in the frame of conventional theory.

Firstly, as Feynman puts it: "We use the term \textit{mass} as a quantitative measure of \textit{inertia} ..."  
 (\cite{Feynman1}, section 9-1).
Therefore, when a scattering particle with mass $M$ is not completely free but \emph{partially bound}
 to other adjacent particles, then the impinging (or probe, e.g.~neutron) particle  scatters on
an  object with \emph{higher} measure of inertia --- or higher mass --- than $M$. This is because 
the adjacent massive particles excert forces on the scattering particle. 
These forces 
may be due to some conventional binding mechanisms
(such as ionic or van der Waals forces; chemi- or physisorption).
However these forces can only hinder the motion of the scattering particle, and thus 
can never cause  an increase of the particle's mobility.
In other terms,  the particle is \emph{dressed} by
certain environmental degrees of freedom, and this dressing  increases  its inertia, or
equivalently, its \textit{effective} mass $M_{eff}$
\begin{equation}
M_{eff} \geq  M \equiv M_{free}.
\label{M-eff}
\end{equation}
This reasoning corresponds to a well understood effect, often observed in scattering
from condensed systems; cf. \cite{Watson,Tietje,Sears}.  The last relation
holds under the conditions of NCS (DINS) and INS too.

This effect can be also shown  by referring to the aforementioned energy conservation relation,
see (\ref{E-conserv2}), here including a  term $E_{int} > 0$ describing the
atom-environment interaction:
\begin{equation}
\bar{E}  =
\frac {\hbar^2 \bar{K}^2}{2M}  + E_{int} 
\label{E-conserv3}
\end{equation}
where $\bar{E}$ and $\bar{K}$ refer to the  center of a peak (measured by a specific detector).
Thus there will be a reduced amount of energy, $\bar{E}- E_{int}$,  available as kinetic energy to  the
recoiling particle.
As pointed out above,  $\bar{E}$ and $\bar{K}$  are determined from the kinematics of the neutron, in contrast
to $E_{int}$,     which is a quantity of the scattering system.

Let us now try to  fulfill this equation with a pair $(E_{IA},K_{IA})$ being
determined by the conventional theory in the impulse approximation (IA), for which holds  
\begin{equation}
E_{IA} = \frac {\hbar^2  K_{IA}^2}{2 M_{eff}} \label{IA-assumption}
\end{equation}
Obviously,  from the last two equations follows   $M_{eff} > M$.

\begin{figure}[ht]
\begin{center}
\includegraphics[width=100 mm]{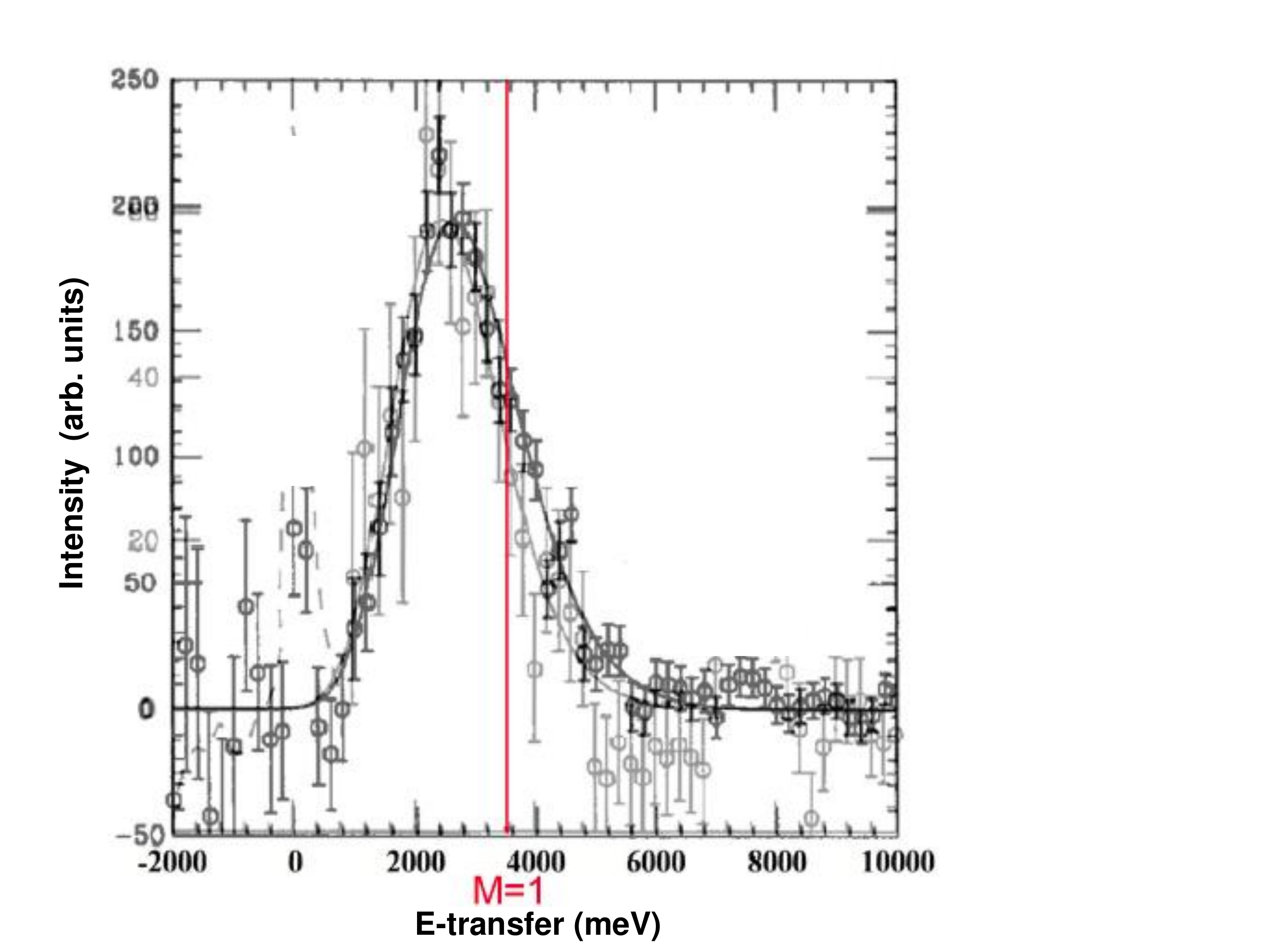}
\end{center}
\caption{\label{SokolFig} The conventional effect $M_{eff} > M$ --- 
DINS from chemisorbed H in
C$_8$K, adapted from figues of \cite{Sokol-White}.
The maxima of the H peaks are about
2620 meV 
The vertical red line  indicates the position of a  $M=1$ a.m.u.~recoil peak (at 3540 meV),
 assuming that the conventional theory (here: the IA)
is obeyed.  --- 
Black points and fitted line: KH,  
Grey points and fitted line: C$_8$K$_{0.9}$H$_{0.13}$.
The spectra were recorded with a selected final neutron energy $E_f =$4280 meV.
The peak maxima correspond to $ M_{eff} = 1.2$ a.m.u.}
\end{figure}

%%%%%%%%%%%%%%%%%%%%%%  SOKOL 
This effect may be  demonstrated with the aid of a DINS result by 
Sokol et al.~\cite{Sokol-White},   obtained 
from H atoms produced by  chemisorbed, dissociated H$_2$ in the graphite intercalation compound (GIC) C$_8$K.
 The experiment was done with a so-called
 \textit{inverse geometry} TOF spectrometer (of Argone Nat.~Lab., USA)
with energy selection in the \textit{final} flight path and chosen momentum trasfer  with $K = 39$ \AA$^{-1}$,
which is high enough for the IA to be valid. Additionally, an accompanying DINS experiment from the ionic solid
KH was done with the same instrumental setup. The similarity between the H recoil peaks in KH and the 
GIC, see Fig.~\ref{SokolFig}, led to the conclusion  that the state of H
is similar in both samples \cite{Sokol-White}. The
measured position of the H peaks corresponds to a very high effective mass, i.e.  
$ M_{eff} = 1.2$ a.m.u.

%%%%%%%%%%%%% SOKOL WHITE 

Thus one obtains the following experimentally testable predictions
of conventional theory:

In specifically designed scattering experiments with  fixed  $K$-transfer  --- the so-called
\textit{constant}-$K$ measurements --- the measured recoil (i.e. kinetic) energy of an atom
with mass $M$  
will be \emph{smaller} than that predicted by the IA.  
This is tantamount to a two-body collision of a probe particle (neutron) 
with a fictitious free atom of mass $M_{eff} >M$. 
                    
For more  references about FSE    and associated discussions of 
"effective mass"   the interested reader may consult Ref.~\cite{Dreismann-Quanta}.

 %%%%%%%%%%%%%%%%%%%%%%% EXPERIMENTAL  
\section{Experiments on the new scattering effect}

In this section, the obtained  theoretical  results are compared with some actually performed  experiments.
The derivations of the preceding section should apply to all neutron scattering subfields of
interest (i.e.~INS, NCS, DINS) as the derivations 
do not contain any specific assumption being valid in one subfield only.
The presented experimental results may be considered as  examples of the 
WV-TSVF-theoretical analysis of Section 5. 

The revised phyiscal intuitions offered by the theoretical result \cite{Aharonov-NJP} 
outlined in Sec.~2 may be understood as
 the reason that led us  to the analysis of the experiments considered here 
 (and several others; see discussion below). 
 In short: 
 The measured scattering signal by a detector is due to a quantum superposition  of the neutron 
\begin{itemize}
 \item[(1)] {colliding 
 with the atom (and thus giving  a "positive" momentum transfer $\hbar K$ to the atom) and} 
 \item[(2)]{
 being transmitted without collision, i.e.~giving a "zero" momentum transfer,} 
 \end{itemize}
with the superposition of both 
 causing the atom to receive an additional  "negative" component $-\Delta (\hbar K)$ 
 to its total momentum.

     %%% FIGURE COWLEY  %%%%%%%%%%%%%%%%%%%%%%%%%%%% 
\begin{figure}[ht]
\begin{center}
\includegraphics[width=100 mm]{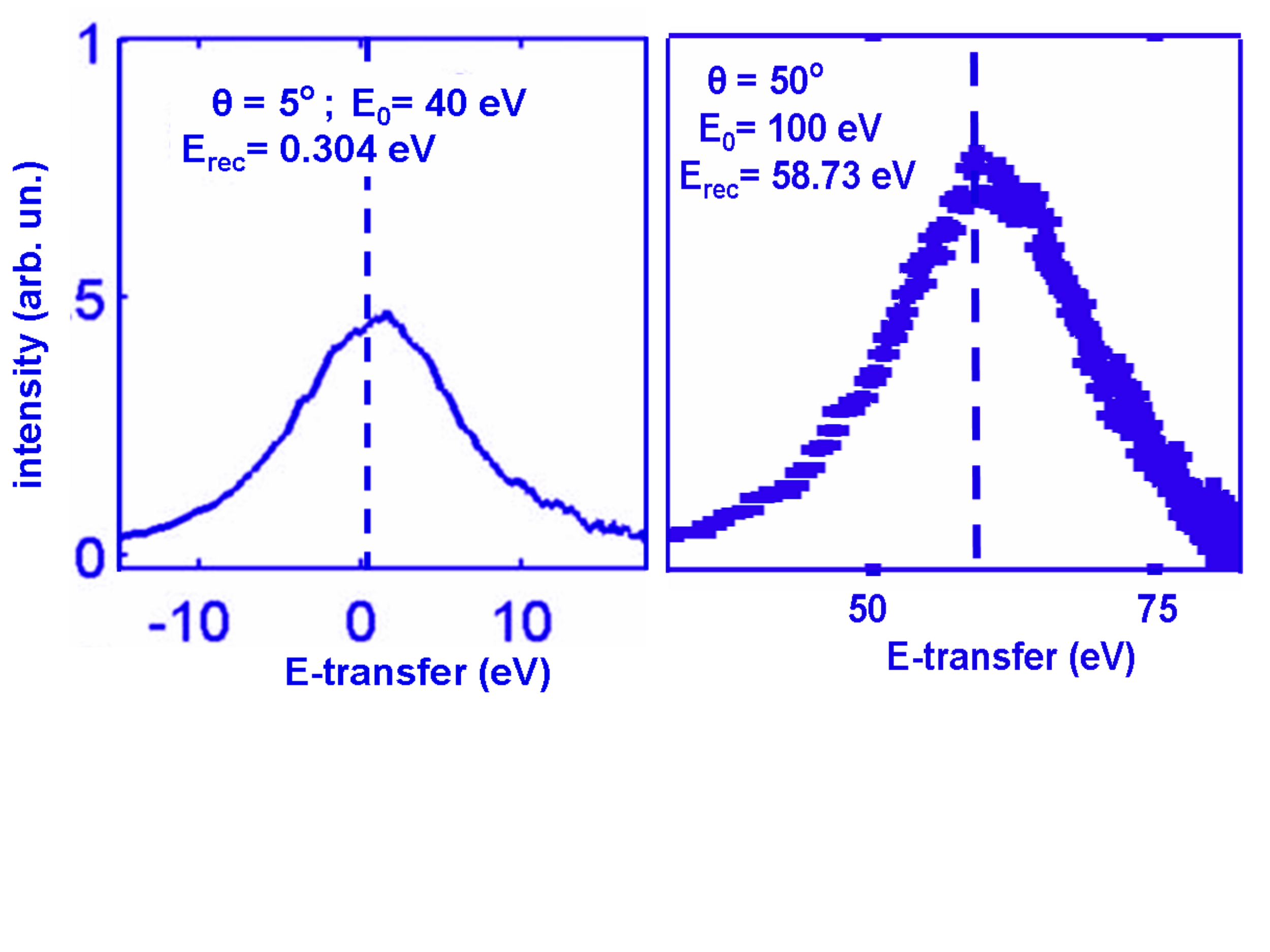}
\end{center}
\caption{\label{Fig-Cowley} Two examples of DINS spectra from a
solid polymer
(polyethylene, $[- \textrm{CH}_2 -]_n $) measured with the TOF
spectro\-me\-ter MARI of ISIS \cite{Cowley2010}. Peak-shifts to \emph{higher} energy
transfers than the conventionally expected recoil energy $E_{rec}$ (vertical lines)
are clearly visible. This effect corresponds to a \emph{lower} effective
mass $M_{eff}(\textrm{H}$) of the recoiling H: 
$ M_{eff}(\textrm{H}) \approx  0.91 - 0.96 $ a.m.u.
See \cite{Cowley2010}  for experimental details and more examples. 
}
\end{figure}

%%%%%%%%%%%%%%%%%%%%%%%%%%%%% Cowley
\subsection{NCS/DINS from H atoms of a solid polymer---MARI experiment}

An experimental demonstration of the new quantum effect under consideration can be found in
the data of \cite{Cowley2010}.  The NCS (or DINS) experiments were carried out by Cowley and
collaborators with the TOF spectrometer  MARI \cite{MARI} of the neutron spallation source ISIS
(Rutherford Appleton Laboratory, UK).
The sample (a foil of a solid polymer, low-density polyethylene) was at room temperature.

   Fig.~\ref{Fig-Cowley} shows two examples of the extensive
measurements reported in \cite{Cowley2010}. The
depicted
recoil  peaks are mainly due to scattering from protons (H atoms), due to the high
incoherent 
scattering cross-section of H \cite{Squires,Watson}.
The vertical  lines show the $E$-transfer values 
according to conventional theory,
Eq.~(\ref{recoil}).  The centroids  of the measured recoil peaks
are markedly shifted to
\textit{higher} energy transfer than conventionally expected.
As discussed in subsection~\ref{EffectiveMass},
this shift is equivalent to a \textit{smaller}
effective mass of the recoiling H atom.

One might object that the shown data contain an additional small contribution from the C recoil.
However this is located at smaller $E$-transfers than that of H, due to their mass difference
\cite{Watson}.
Therefore the preceding qualitative conclusion remains unaffected.
 
It may be noted that the shown NCS peaks are very broad and asymmetric,
which is due to the (very)
low resolution of the employed modified setup  of MARI \cite{Cowley2010}. This makes a
quantitative analysis to determine the peak-position impossible. Nevertheless, visual inspection
of the data shows that the centroids of the peaks are \emph{shifted to higher energy} roughly by
5-10 \% of the recoil energy,
which equivalently  means that  the effective mass
$M_{eff}$ of the recoiling H atoms is \emph{smaller} than $M_{H}=1.0079$ a.m.u. of a
free H
by the same percentage
\begin{equation}
M_{eff}(\textrm{H}) \approx  0.91 - 0.96 \ \ \textrm{a.m.u.}
\label{Meff-H}
\end{equation}

    \begin{figure}[ht]
\begin{center}
\includegraphics[width=90 mm]{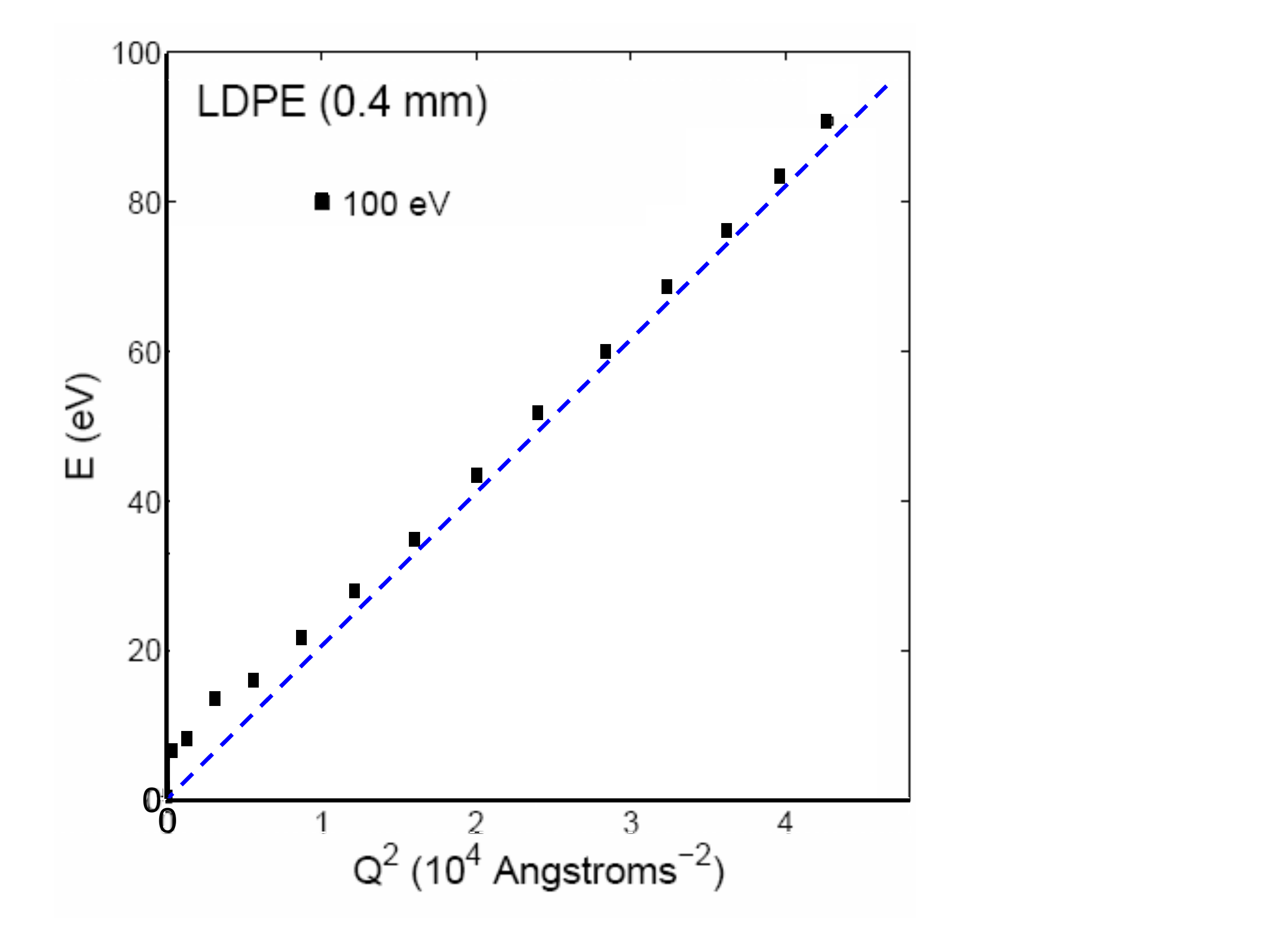}
\end{center}
\caption{ \label{Cowley-Fig-all} Data from NCS (DINS) from a
solid polymer
(low-density polyethylene, LDPE), $[- \textrm{CH}_2 -]_n $) measured with the TOF
spectro\-me\-ter MARI of ISIS with excitation energy $E_0 = 100$ eV. 
\cite{Cowley2010}. The straight line corresponds to the recoil parabola of
conventional theory. All data points (center-of-gravity of recoil peaks) 
show the effect under consideration.
i.e. an apparent reduced effective mass of the scattering H-atom. 
For experimental details and more examples, see \cite{Cowley2010}.
(Data taken from Fig.~12 of \cite{Cowley2010}.)    }
   \end{figure}

Throughout this section we presumed that the
\emph{calibration} of  MARI as employed in the experiment
\cite{Cowley2010} was correct, i.e.~that the applied TOF method (see
Sec.~5)  used correct instrumental parameters.

Moreover, it seems important to look at the additional (about 50) DINS-results being reported
in Fig.~12 of Ref.~\cite{Cowley2010}, which were obtained at three different
excitation energies ($E_i = 20, 40,100$ eV).  Remarkably, \textit{all} those H-recoil peaks 
 --- i.e. with no exception! --- appeared to show \textit{positive}
$E$-transfers;
cf.~Fig.~\ref{Cowley-Fig-all} which shows data obtained with neutron's initial 
energy $E_i =100$ eV. This observation
underlines considerably the  reliability of the considered  "anomalous" $E$-transfer shift.
%%%%%%%%%%%%%%%%%%%%%%%%%%%%%%%%%%%%%%%%%%%%%%%%%#########
However, the main aim of that investigation \cite{Cowley2010}
was  motivated differently  --- i.e.,  the focus was the measurment of  the cross-section of 
H,   and to show a weakness of the "inverse geometry" TOF-spectrometer VESUVIO  of ISIS --- 
this striking experimental finding  remained fully
unnoticed and/or uncommented in the discussions of Ref.~\cite{Cowley2010}.

%%%%%%%%%%%%%%%%%%%%%%%%    NANOTUBES   %%%%%%%%%%%%%%%%%%%%%%%%%%%%%%%%%%%%%%%%%%%%%%%%%%%%%%%%%%%%%%%%%%%%
\subsection{\label{H2nanotubes} INS from single H$_2$ molecules in carbon nanotubes}

Another surprising result from incoherent inelastic neutron scattering was observed by Olsen et
al.~\cite{Olsen-H2} in the quantum excitation
spectrum of H$_2$  adsorbed in multi-walled nanoporous carbon (with
pore diameter about \mbox{8--20 \AA}).

The INS (or IINS) experiments were carried out at the new generation
TOF spectrometer ARCS
of Spallation Neutron Source SNS (Oak Ridge Nat.~Lab., USA) \cite{ARCS}. In this
experiment, the temperature was $T= 23$ K, and the incident neutron
energy $E_i$ was 90 meV.   The latter implies that the energy
transfer cannot excite molecular vibrations (and thus cannot break the molecular
bond), but only excite rotation and translation (also called recoil)
of H$_2$ which interacts only weakly with the substrate:
\begin{equation}
E = E_{rot} + E_{trans} 
\label{rot-trans}
\end{equation}

\begin{figure}[t]
\begin{center}
\includegraphics[width=90 mm]{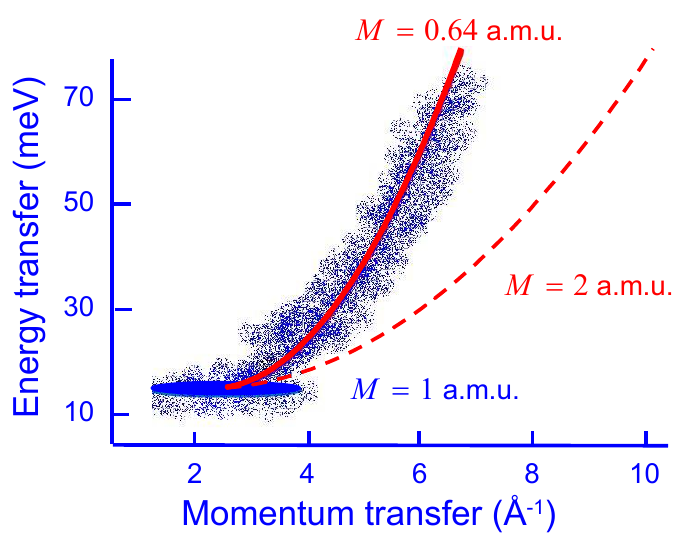}
%{Fig4.pdf}
\end{center}
%%%%%%%%%%%%%%%%%%%%%%%%
\label{Fig4}
\caption{Schematic representation of incoherent inelastic neutron scattering results
from H$_2$ in carbon nanotubes, with incident
neutron energy $E_0 =90$ meV; cf.~Fig.1 of Ref.~\cite{Olsen-H2}. The translation motion of the
 recoiling H$_2$ molecules causes  the observed continuum of
 intensity, usually called ``roto-recoil" (white-blue ribbon),  starting at the well visible first rotational excitation
 of H$_2$  being centered at
$E \approx 14.7$ meV and $K \approx 2.7$ \AA$^{-1}$ (blue ellipsoid).
The $K-E$ position of the latter is in agreement with conventional theory.
 In contrast,
 a detailed fit (red parabola; full line) to the roto-recoil data reveals a strong reduction of the
 effective mass of recoiling H$_2$, which appears to be only 0.64 a.m.u. (The red broken line, right parabola) represents
the conventional-theoretical parabola with effective mass 2 a.m.u.)
 For details of data analysis, see  \cite{Olsen-H2}. (Reproduced from Ref.~\cite{Dreismann-Quanta}, Fig.~4,  with permission from \textit{Quanta}.)
 }
\end{figure}

The experimental two-dimensional incoherent inelastic neutron scattering intensity map $S(K,E)$ of H
(after background subtraction) is shown in Fig.~\ref{Fig4},
which is taken from the original paper \cite{Olsen-H2}. The
following features are clearly visible.
First, the  intensive peak  centered at
$E_{rot} \approx 14.7$ meV  is due to  the well known first rotational excitation
$J=0 \rightarrow 1$
of the H$_2$ molecule \cite{Mitchell-Buch}.
Furthermore, the wave vector transfer of this peak is  $K_{rot} \approx 2.7$ \AA$^{-1}$.
Thus the peak position in the $K$--$E$~plane shows  that the experimentally determined
mass of H   that  fulfills the relation $E_{rot} =(\hbar
K_{rot})^2/2M_{H} $ is (within experimental error) the mass $M_{H}$
of the free H atom, 
\begin{equation}
\textrm{rotation:}\ \ \ \ M_{H} = 1.0079\ \  \textrm{a.m.u.}
\label{M-rot}
\end{equation}
namely, ${M}_{eff}(H) = M_H$.  (a.m.u.: atomic mass units.)
In other words, the location of this rotational excitation in the $K$--$E$~plane agrees with
conventional  theoretical expectations for IINS,
according to which each  neutron scatters  from a single H \cite{Mitchell-Buch}.
Recall that an agreement with conventional theory was also observed  in the case of scattering
from $^4$He \cite{Glyde}; see Fig.~\ref{He-recoil}.

%%%%%%%%%%%%%%%%%%%%%%%%%%%%%%%%%%%%%%%%%%%%%%%%%%%

Moreover,
the authors provide  a detailed  analysis of
the  roto-recoil data from incoherent inelastic neutron scattering, as shown in Fig.~\ref{Fig4}, 
and extracted a
strongly reduced   effective mass of the whole recoiling H$_2$ molecule (left parabola, green line);
see Eq.~(\ref{recoil}):
\begin{equation}
\textrm{translation (recoil):}\ \ \ \   {M}_{eff}(\textrm{H}_2) \approx 0.64  \pm 0.07 \ \ \textrm{a.m.u.}
\label{M-recoil}
\end{equation}
This is in blatant contrast to the conventionally expected value $M(\textrm{H}_2) =2.01$ a.m.u. for a
freely recoiling  H$_2$ molecule (right parabola, red line). (Recall that the neutron-molecule
collision does not break the molecular H-H bond.)

An extensive numerical  analysis of the data is presented  in
\cite{Olsen-H2}, being based on
time-of-flight data analysis (cf.~Section~\ref{sec:4}) and the analysis of the
measured data within conventional theory
\cite{Squires,Mitchell-Buch}.

This strong reduction of effective mass, which is far beyond any
conceivable experimental error, corresponds to a strong reduction of
momentum transfer by the factor $~ 0.566$. Namely, the observed
momentum transfer deficit is about $ - 43\%$  of the conventionally
expected momentum transfer. This  provides first experimental
evidence of the new \emph{anomalous} effect of  momentum-transfer deficit
in an elementary neutron collision with a recoiling molecule.

Recall that, as explained above (see Subsec.~\ref{EffectiveMass}), every H$_2$-substrate  conventional 
binding 
must increase
the molecule's effective mass. Thus these findings from IINS are in clear contrast to every conventional
(classical or quantum)  theoretical 
expectation. However, they have a natural (albeit qualitative, at present) interpretation in the
frame of modern theory of WV and TSVF.

Incidentally, it may be noted that  the same \emph{calibration} of ARCS was used in
both experiments \cite{Glyde} and \cite{Olsen-H2}.

Let us now make a comparison with a related 1-$dimensional$ experiment, which  may shed more light into
the observations under consideration.

The  above experimental results  also show that the two-dimensional spectroscopic technique,
as offered by the  advanced TOF-spectrometer ARCS,  represents a
powerful method that  provides novel insights into quantum dynamics of molecules and condensed
matter. Clearly, this is due to the fact that  $K$ and $E$ transfers can be measured over a broad
region of the $K$--$E$~plane. This advantage makes these new instruments superior to the common
one-dimensional ones (like TOSCA at ISIS spallation source, UK), in which the detectors can only measure  along a
single specific  trajectory in the $K$--$E$~plane. (TOSCA measures along two such trajectories
\cite{TOSCA}.)

As an example, consider the results of
\cite{Georgiev2005} from molecular H$_2$ adsorbed in single-wall carbon nanotubes   (which is
similar to the material of \cite{Olsen-H2}) at $T \approx 20$ K, as investigated with TOSCA. 
Also this paper
reports the measurement of the roto-recoil spectrum, but as a function of $E$ only (due to the aforementioned
single trajectory in the $K$--$E$~plane being instrumentally accessible). Therefore for the theoretical
analysis of the data the mass of H$_2$ was
\emph{fixed}  to its conventionally expected value of 2 a.m.u., and thus the
strong anomalous effect,  Eq.~(\ref{M-recoil}), remained unnoticed; 
see \cite{Georgiev2005}, p.~903.

%%%%%%%%%%%%%%%%%%%%%%%%%%%%%%%%%%%   DICUSSION   
\section{Discussion}

Due to the novelty of the theory of WV and TSVF, several counter-intuitive features 
therein and in its applications may appear
"strange" or even "wrong".   Aiming to clarify some of them, here we provide    
additional remarks and explanations. \\

$(A)$  \textit{Instrumental calibration} --- The results presented above have  considerable consequences for the
calibration of the associated TOF spectrometers. In particular, in  NCS (DINS) instrumentation, it is a
common practice  to use the recoil peaks of light atoms (typically H, D or He) to achieve an 
"optimal" calibration
 of the spectrometer. In other words,  the
measured positions of a peak in the $K$--$E$~plane --- together with the standard free-atom recoil expression of
Eq.~(\ref{recoil}), sometimes also improved with FSE --- 
are used in order  to  "fine
tune" the numerical values of the instrument's parameters determining the measured TOF values.    
 Obviously, such a "optimization" of the calibration leads automatically
to an  artificial  agreement of the data with  conventional theory. \\

$(B)$ Note that in in our approach, the two operators $\hat{q}$ and $\hat{P}$ occurring in the von
Neumann-type interaction Hamiltonian of Eq.~(\ref{H-int}) refer to two \emph{different} quantum
systems. Thus, according to Vaidman \cite{VaidmanComment2014}, the concept of WV  arises
here  due
 to the
interference of a quantum \emph{entangled} wave and  therefore  it has no analog in classical wave
interference. This further supports
the conclusion   that WV  is a genuinely quantum concept and not some kind of "approximation".

  In contrast, conventional neutron scattering theory treats the neutron as a classical mass point. 
Namely, consider the basic formulas of the theory, e.g.~the expression for the partial differential cross-section
\begin{equation}
 \frac{d^2\sigma_A}{d\Omega d\omega}  =
  \frac{k_f}{k_i} b_A^2 S(\textbf{K},\omega)
     \label{pdcs}
\end{equation}
($E=\hbar \omega$, $d\Omega$: solid angle measured/covered by the detector.)
$S(\textbf{K},\omega)$ is the dynamic structure factor of the scattering system, which contains
degrees-of-freedom of the scattering particles only. 
This equation contains  \textit{no dynamical} variable of the neutron at all --- the neutron is 
"degraded" to a classical object and only its scattering length $b_A$ --- a $c$-number 
capturing its scattering properties from atom $A$ --- appears here.     
This fact is a specific feature of the first Born approximation and/or first-order perturbation
theory  which the conventional theory is based on  \cite{Squires,Watson}.\\

$(C)$  The momentum-transfer deficit, and the associated effective-mass reduction of the scattering particle $A$,
may appear to someone as violating the energy and momentum conservation laws of basic physics. However,
this is not the case, because the scatterer $A$ is $not$ an isolated, but an \textit{open} quantum system.    
Thus we may say that the quantum dynamics of the "environment" of $A$, which participates to the 
neutron-$A$ scattering, is indispensable  for the new WV-TSVF effect  under consideration. Thus it may
be helpful to     write down the "conservation" relations
\begin{equation}
E_{H+env}= - E_n \ \ \ \  \textrm{and} \ \ \ \  \hbar K_{H+env} = - \hbar K_n
\label{H+env}
\end{equation}
which express energy and momentum conservation for the case that the
\textit{environment} (indicated with the subscript \textit{"env"}) of the scattering H is not neglected.
The theoretically derived momentum transfer deficit, when interpreted (or better: misinterpreted)  
in terms of conventional theory,
is equivalent to scattering of a neutron from  a ficticious particle representing the \textit{whole} 
H+environment system, which has a mass $M_{eff}$ being smaller than the mass of a free H-atom --- 
or, in other terms, the environment of H seems to have a negative mass! \\
  
$(D)$  The INS results from H$_2$ in C-nanotubes \cite{Olsen-H2} discussed in subsection \ref{H2nanotubes} 
appear contradictory --- in the frame of conventional theory --- because of the following two main points:

$(i)$ The observed  $J=0 \rightarrow 1$  rotational excitation of the H$_2$ molecule by
the collision with a neutron exhibits a $ M_{eff}\approx 1$ a.m.u., see Eq.~(\ref{M-rot}), as
conventionally expected. Namely, the scattering is incoherent, i.e.~the neutron exchanges 
energy and momentum with one H only.
   
$(ii)$ In contrast, in the same experiment,  the $M_{eff}$ of the observed roto-recoil response
of the whole H$_2$ molecule is not 2 a.m.u.~as it should (because the whole H$_2$ 
undergoes a translational motion), but only $M_{eff}\approx 0.64$ a.m.u.,   see Eq.~(\ref{M-recoil}).
  
However, this "contradiction" just disappears in the light of the new theory, because it simply
implies that the quantum \textit{environment} of H in case $(i)$ must be \textit{different} 
from that in case  $(ii)$ --- the locally rotating  H$_2$ is much less influenced by its environmental 
interactions than the translating  H$_2$, which necessarily interacts with a greater environmental part
during its translational motion caused by its collision with the neutron.  \\
  
$(E)$  In the scientific literature one often meets the criticism that \textit{post-selection} 
just means
"throwing  out some data". In  the experimental context at issue,
however,   post-selection 
 certainly means "performing a concrete measurement on the system,
using a well defined detector,  and analyzing the  measured data only".
In this context it may be helpful to mention the paper by Wu \cite{Wu}, in which he
offers a slight extension of the WV and TSVF that does not discard any data. Furthermore, it
is shown how the matrix elements of   any generalized state (pure or mixed) can be
directly read from appropriate weak measurements.\\

$(F)$  Concerning the applicability of WV (and WM) to the neutron scattering process, one
may object that the neutron-nucleus potential (i.e., the conventional Fermi pseudopotential
\cite{Squires}) may be not "weak" in the specific sense of the new theory. Therefore
it is helpful to mention the recent generalization
of the considered WM and WV theory by
Oreshkov and Brun  \cite{Oreshkov}. The   authors showed
that WM are universal, in the sense that
any generalized measurement can be decomposed into a sequence of WMs. This important
theoretical result is further supported by the work of Qin et al. \cite{{Qin2015}}, who
showed that the main WM results can be extended to the realm of \emph{arbitrary}
measurement strength.\\

$(G)$ Very recently, some qualitatively new theoretical results of  WV-TSVF  have been
obtained, a few of which should be mentioned here: 

$(a)$ The quantum Cheshire-Cat effect \cite{Cheshire2013}, which has also been observed experimentally
 \cite{Cheshire2014}.  
 
$(b)$ The predicted quantum violation of the Pigeonhole principle \cite{pigeonhole}.
  
These new results point once more to a very interesting structure of quantum mechanics that was hitherto unnoticed. 
Moreover, they also shed new light on the very notions of separability, quantum correlations and   
quantum entanglement \cite{Horodecki}. \\
     
$(H)$ The exploration of the "anomalous" momentum exchange taking place on a mirror of a MZI \cite{Aharonov-NJP}, 
as discussed in Sec.~\ref{NJPpaper},
indicates that this effect could also be relevant in the context of the widely applied methods 
of small-angle neutron scattering (SANS), small-angle X-ray scattering (SAXS) and reflectometry on surfaces and/or 
thin multilayer  structures of physico-chemical and biological materials. \\
              
$(I)$ Recently, quantum interference and entanglement \cite{Horodecki} effects have been recognized to be
crucial for quantum computing   and quantum information theory; cf.~\cite{NielsenChuang}. 
Hence  it seems interesting to explore the potential applicability of the new quantum features connected 
with WV, TSVF and the  momentum transfer deficit (or energy transfer surplus) to  recently emerging fields of 
quantum computational complexity theory \cite{AroraBarak,Fortnow},  and/or ICT (information and
communication technology).   \\   
							
$(J)$ In view of the experimentally detected effects of Sec.~6, and in particular of the
striking reduction of effective mass due to quantum interference, it seems appropriate to
mention here some speculative ideas concerning a possible 
\textit{practical} and/or \textit{technological} importance of the new theory.
Namely: 

(1) a more mobile (i.e. with smaller effective mass) H atom (or H$^+$ ion) in a properly designed 
fuel cell material 
would facilitate proton transfer and thus also increase the efficiency of the fuel cell;
  
(2) a properly designed solid environment (or nanostructure) of  Li$^+$ ions in a Li-ion-battery 
would result in a lower Li$^+$ effective mass, thus allowing a faster ionic mobility and
a \textit{faster recharging} process --- with obvious advantages for current technological
efforts. \\

Concluding, the present author believes that the theoretical formalism of WM, WV, and 
TSVF  not only sheds new light on interpretational issues concerning fundamental quantum
theory (like e.g. the time-inversion invariance of the basic physical laws, the meaning of 
quantum entanglement and correlations, etc.) but  it also offers a fascinating new guide 
for our intuition to predict new effects, and  also  helps to plan and carry out new
experiments and reveal novel quantum phenomena.

\ack
I wish to thank Erik B.~Karlsson (Uppsala) and Ingmari Tietje (CERN)  for  helpful   discussions,
and COST Action MP1403 -- Nanoscale Quantum Optics -- for partial financial support.\\

 %--------------------------------------------------------------------------------------------

\end{document}